\documentclass[aip,apl,graphic,reprint]{revtex4-2}

\usepackage{graphicx}% Include figure files
\usepackage{dcolumn}% Align table columns on decimal point
\usepackage{bm}% bold math
\usepackage[utf8]{inputenc}
\usepackage[T1]{fontenc}
\usepackage{mathptmx}
\usepackage{etoolbox}
\usepackage{setspace}
\usepackage{xcolor}
\begin{document}

\title{Controlling Skyrmion Lattice Orientation with Local Magnetic Field Gradients}

\author{Duc Minh Tran}
\affiliation{Institute of Physics, Johannes Gutenberg University Mainz, Staudingerweg 7, 55128 Mainz, Germany}
\author{Edoardo Mangini}
\affiliation{Institute of Physics, Johannes Gutenberg University Mainz, Staudingerweg 7, 55128 Mainz, Germany}
\author{Elizabeth M. Jefremovas}
\thanks{Current address: Department of Physics and Materials Science, University of Luxembourg, 162A Avenue de la Faiencerie, L-1511 Luxembourg, Grand Duchy of Luxembourg}
\affiliation{Institute of Physics, Johannes Gutenberg University Mainz, Staudingerweg 7, 55128 Mainz, Germany}
\author{Fabian Kammerbauer}
\affiliation{Institute of Physics, Johannes Gutenberg University Mainz, Staudingerweg 7, 55128 Mainz, Germany}
\author{Dennis Meier}
\affiliation{Department of Materials Science and Engineering, Norwegian University of Science and Technology (NTNU), Sem S{\ae}lands vei 12, 7491 Trondheim, Norway }
\affiliation{Centre for Quantum Spintronics, Norwegian University of Science and Technology (NTNU), 7491 Trondheim, Norway}
\affiliation{Faculty of Physics, University of Duisburg-Essen, Lotharstr. 1, 45057 Duisburg, Germany}
\author{Robert Fr{\"o}mter}
\affiliation{Institute of Physics, Johannes Gutenberg University Mainz, Staudingerweg 7, 55128 Mainz, Germany}
\author{Mathias Kl{\"a}ui*}
\thanks{Email: klaeui@uni-mainz.de}
\affiliation{Institute of Physics, Johannes Gutenberg University Mainz, Staudingerweg 7, 55128 Mainz, Germany}
\affiliation{Centre for Quantum Spintronics, Norwegian University of Science and Technology (NTNU), 7491 Trondheim, Norway}

\begin{abstract}
Precise control over the formation and arrangement of magnetic skyrmion lattices is essential for understanding their emergent behavior and advancing their integration into spintronic and magnonic devices. Using single-pass magnetic force microscopy (MFM), we establish a protocol to nucleate and manipulate skyrmion lattices in soft magnetic CoFeB. By tuning the scan-line spacing to match the intrinsic stripe domain periodicity, the stray field gradient from the MFM tip induces reversible transitions from stripe domains to isolated skyrmions and locally ordered lattices. The resulting skyrmion positions are extracted to compute the local orientational order parameter $\psi_6$, enabling quantitative evaluation of lattice ordering. A systematic improvement in $\langle|\psi_6|\rangle$ is observed with repeated scanning, indicating a transition from a disordered state to ordered hexagonal arrangements. Furthermore, we demonstrate that the lattice orientation can be directly rotated by changing the scanning direction, as confirmed through real-space analysis and fast Fourier transformations. This method enables the local creation, reordering, and deletion of metastable skyrmions on demand, providing unprecedented control over lattice symmetry, order, and orientation.
\end{abstract}


\maketitle

Magnetic skyrmions are topologically stabilized chiral spin textures that exhibit quasi-particle-like behavior, making them promising candidates for spintronic and neuromorphic applications\cite{fert2013skyrmions,jiang2017skyrmions,mühlbauer2009skyrmion,liu2016skyrmions,fert2017magnetic,finocchio2016magnetic,everschor2018perspective,raab2022brownian,song2020skyrmion,grollier2020neuromorphic}. In the low-pinning regime with a flat energy landscape, skyrmions assemble into dense hexagonal arrangements, constituting a fruitful model to study the properties of two-dimensional (2D) ordering phenomena\cite{huang2020melting,meisenheimer2023ordering,zazvorka2020skyrmion,yu2010real,zhang2022room,zhang2024spin}. Information about local correlations and orientations, ordering transitions, and formation of topological defects can be extracted from the dynamics of skyrmion lattices\cite{huang2020melting,gruber2025real,kapfer2015two}. For magnonic interactions, these skyrmionic crystals are of particular interest as complex spin-wave modes\cite{mochizuki2012spin,li2022interaction}, minibands\cite{mohanta2020signatures}, Landau levels\cite{weber2022topological}, edge states\cite{diaz2019topological,diaz2020chiral}, and resonances\cite{okamura2013microwave,satywali2021microwave} can be excited depending on the order and relative alignment of the underlying skyrmion lattice.

While thermally activated skyrmions in a suitable confinement can form lattices with long-range order when assisted with periodic field oscillations to reduce the effect of pinning potentials\cite{gruber2022skyrmion,reichhardt2022statics,gruber2023300,gruber2025real,brems2025realizing}, their diffusive nature eventually promotes melting of the lattice and diminishing correlation. Magnetic textures in multi-repetition stacks with enhanced thermal stability provide a more robust model in sustaining hexagonal arrangements. As the skyrmion size decreases rapidly with the number of repetitions\cite{jefremovas2025role}, magnetic force microscopy (MFM) is usually employed as an accessible in-house technique to study these systems. In this approach, MFM offers a unique opportunity to both image and directly manipulate magnetic textures via the stray field at the tip's apex, enabling control at the local scale. By scanning the magnetic tip in close proximity to the sample surface, a sufficiently strong field gradient is generated that induces transitions from the energetically favorable stripe domain to the metastable skyrmion state\cite{ognev2020magnetic,zhang2018direct,casiraghi2019individual}. However, most conventional MFM implementations use a two-pass imaging scheme, in which the magnetic signal is acquired during a lift scan following an initial topography pass that operates much closer to the surface, i.e., in the so-called tapping mode. This first pass often causes strong tip-induced magnetic perturbations, making MFM an inherently intrusive technique. Consequently, the phase information in the subsequent MFM image reflects an already altered state due to the magnetic perturbation from the preceding surface mapping. While this approach can work reasonably well in hard magnetic systems that can withstand such disturbances\cite{lee2020writing, zelent2021skyrmion}, soft magnetic materials like CoFeB are far more susceptible, and the tip can readily distort or even erase magnetic textures during the topography scan\cite{casiraghi2019individual}.

Here, we employ single-pass MFM to overcome these limitations by decoupling topography and magnetic sensing, enabling minimally invasive imaging and controlled tip-induced manipulation. Adjusting the spacing $\Delta$ between adjacent scan lines allows for deterministic and fully reversible nucleation and annihilation of skyrmions. In the “writing mode”, where the tip is scanned close to the sample surface, stripe domains are locally converted into skyrmions. The resulting textures and lattices are subsequently captured in the minimally invasive “imaging mode” at a fixed, extended lift height for quantitative ordering evaluation. We further demonstrate that individual skyrmions can be selectively displaced to improve lattice order, enabling a transition from a disordered ensemble to a long-range ordered hexagonal packing. Finally, we show that the lattice orientation can be directly rotated by changing the writing direction without creating new or deleting existing skyrmions. This ability to freely reconfigure local lattice ordering and orientation has not been demonstrated previously and provides a versatile platform for exploring lattice-dependent phenomena, including anisotropic spin-wave propagation.

\begin{figure*}
\includegraphics{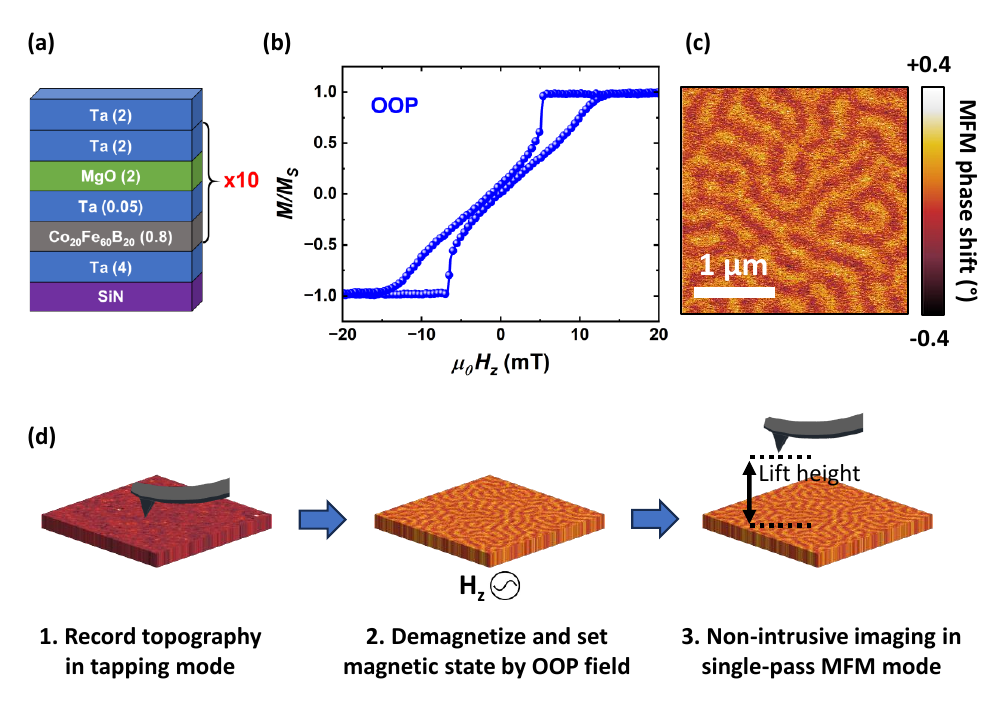}
\caption{\label{fig:wide} Skyrmion-hosting stack and measurement scheme. (a) Schematic of the ten-repetition ferromagnetic stack. (b) Normalized OOP SQUID magnetometry loop of the continuous multilayers at 300 K. (c) Demagnetized magnetic domain pattern recorded with single-pass MFM. (d) Single-pass procedure to obtain MFM measurements with minimized perturbation by the stray field of the tip.}
\end{figure*}

Figure 1a details the magnetic multilayer stack of Ta(4)/[Co\textsubscript{20}Fe\textsubscript{60}B\textsubscript{20}(0.8)/Ta(0.05)/MgO(2)/Ta(2)]\textsubscript{x10}, deposited by DC/RF magnetron sputtering using a Singulus Rotaris tool with a base pressure of 3×10\textsuperscript{-8} mbar on a SiN substrate. Layer thicknesses are given in nanometers in parentheses. The Ta/CoFeB interface provides the interfacial Dzyaloshinskii–Moriya interaction (DMI) to stabilize chiral magnetic textures, while the CoFeB/MgO interface generates a sizeable perpendicular magnetic anisotropy (PMA)\cite{dieny2017perpendicular}, as seen in Figure 1b. A thin 0.05-nm Ta dusting layer is introduced between the CoFeB and MgO to balance competing energy terms from anisotropy, exchange, dipolar, and DMI to host skyrmions, and also to optimize the energy landscape for skyrmion lattice formation by low pinning dynamics\cite{gruber2022skyrmion,zazvorka2019thermal}. The core skyrmion hosting configuration of Ta/CoFeB/MgO is repeated 10 times to promote dipolar stabilized skyrmions\cite{jefremovas2025role}.
SQUID magnetometry confirms that the magnetic moment of our system saturates at approximately 18 mT in an out-of-plane magnetic field, which limits the use of most conventional MFM probes operating in the two-pass scheme. To avoid significant perturbation, we imaged the magnetic textures using single-pass MFM with an ND-MDT Ntegra Prima scanning probe microscope equipped with a built-in electromagnet. Commercial MESP-LM-V2 magnetic probes (Bruker) were used at a fixed lift height of 200 nm, producing a nominal stray field below 1 mT acting on the sample in the imaging sequences\cite{zhang2018direct,casiraghi2019individual,liu2025room,corte2020comparison}. In tapping mode, these probes generate stray fields exceeding the sample’s saturation field, sufficiently enabling nucleation of dense skyrmion lattices\cite{ognev2020magnetic}.
The single-pass mode decouples the topographic scan from the magnetic phase detection, bypassing the need for consecutive surface mappings. This approach effectively eliminates topography-related interference, enabling minimally invasive imaging of magnetic textures. As shown in Figure 1c, a maze domain pattern is observed in the demagnetized state, consistent with systems exhibiting PMA and low remanence. Dark (bright) contrast represents magnetization pointing along the $-z$ ($+z$) direction. The undistorted stripe configuration and equal distribution of the up and down magnetization confirm minimal perturbation of the single-pass technique. Our measurement procedure is detailed in Figure 1d. First, we recorded the topography of the entire region of interest in conventional tapping mode. Subsequently, the sample was demagnetized to reset any influence from the tip in the previous topography scan. Finally, non-contact scans were performed to collect the MFM phase signal at a fixed lift height, using the initially recorded topography as surface mapping reference. 

We show that by tuning the probe's stray-field intensity acting on the sample via the lift height, magnetic domains can shrink or expand depending on their relative polarity to the probe's magnetization. Operating in tapping mode, where the tip oscillates in the regime of 1 - 10 nm, induces a sufficient stray field to saturate local domains beneath the probe\cite{casiraghi2019individual,ognev2020magnetic,corte2020comparison}. This strategy allows us to create skyrmions locally by cutting long stripe domains whose magnetization is antiparallel to the one of the tip. Figure 2a demonstrates the local and reversible transition from the energetically favorable stripes to skyrmions, triggered by the scanning magnetic field gradient. All MFM images in this work were obtained with the tip magnetized in the $+z$ direction, antiparallel to the magnetization of the skyrmion cores. Minimally invasive imaging scans were performed at a lift height of 200 nm above the sample surface, whereas manipulations (writing and erasing) were carried out consistently in tapping mode at a low scan speed of 5 $\mu$m/s to sufficiently convert stripes into skyrmions and reduce the influence of other scanning parameters. The line spacing $\Delta$ was controlled by adjusting the number of scan lines constituting each complete image. A small OOP field of 0.25 mT was applied to bias the sample's magnetic state. This external field offsets the system from the stable ground state and slightly favors one magnetization at the expense of the other, thereby facilitating an efficient stripe-to-skyrmion transition when a writing pass is conducted. In addition, the field was kept low to promote a dense domain configuration over isolated stripes and skyrmions. Five writing passes, each consisting of 16 scan lines with spacing $\Delta$ $\approx$ 312 nm, were performed along the 0$^{\circ}$ direction to slice stripe domains into skyrmions, as illustrated in Figure 2b. The line spacing $\Delta$ is crucial to facilitate efficient stripe-to-skyrmion conversion and dense lattice formation, and was chosen to closely match the domain periodicity $L$ $\approx$ 320 nm (see Supplementary Material S2). Insets present the Fourier transformation (FFT) of the phase images, starting with a diffuse ring pattern, characteristic of the maze domain pattern, to periodic peaks with sixfold rotational symmetry after skyrmion lattices have been written, and finally a diffuse ring pattern again after they have been erased, confirming the reversible process. The nucleated skyrmions are aligned along the fast scan axis, indicating significant tip-induced repulsion affecting and ordering the low-pinning skyrmions in our system. A single erasing scan was performed with $\Delta$ $\approx$ 10 nm—too narrow to split any domains, but sufficiently dense to push them out of the scanning path, as illustrated in Figure 2c. In the densely populated domain environment, displaced skyrmions had no available space to occupy and were consequently annihilated by rejoining back to stripe domains. This process effectively erases the previously nucleated skyrmion lattices, reverting the system to its energetically favorable stripe domain state.

\begin{figure}
\includegraphics{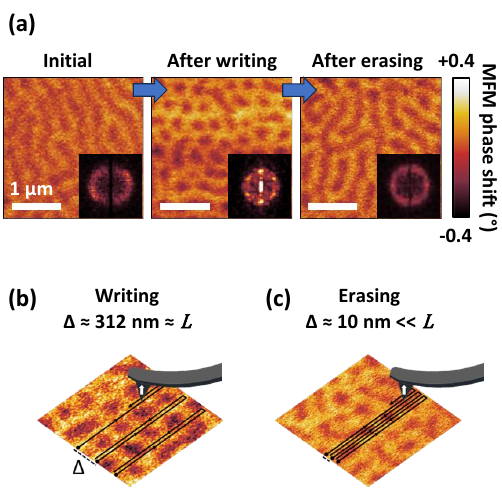}
\caption{\label{fig:epsart2} Local nucleation and annihilation of magnetic skyrmions by the MFM tip. (a) Transition from the initial maze domain state to a skyrmion state after 5 writing scans, and erasing back to maze domains, driven by scanning the sample in tapping mode with different line spacings $\Delta$. Insets show the corresponding FFTs of the MFM images. Schematic of the process to (b) write skyrmions using $\Delta$ $\approx$ 312 nm, closely matched to the domain periodicity, and (c) erase skyrmions by driving skyrmions into annihilation with $\Delta$ $\approx$ 10 nm.}
\end{figure}

Nucleating local skyrmion lattices with MFM has been demonstrated in various ferromagnetic systems\cite{ognev2020magnetic,jin2025local,zhang2018direct,zhang2022room}. However, when conventional two-pass MFM is used, topographic scanning inevitably perturbs the sample before any magnetic phase information can be collected. This hinders the ability to freely manipulate magnetic textures into complex arrangements. Consequently, most studies have focused on the spontaneous nucleation of skyrmions, typically exhibiting lower lattice order and mostly short-range correlations, even in material systems designed to host skyrmion lattices (see Supplementary Material S3). In contrast, single-pass MFM can directly access the positions of individual skyrmions, allowing for their controlled rearrangement and the realization of 2D transitions with progressively enhanced lattice order, which has so far only been studied at a global scale\cite{gruber2025real,meisenheimer2023ordering,huang2020melting,zazvorka2020skyrmion,zhang2024spin}. Figure 3a illustrates the repulsive interaction generated at the MFM tip’s apex, which acts on local skyrmions with anti-parallel magnetization in a low-pinning system\cite{casiraghi2019individual,ognev2020magnetic}. At low magnetic fields, the system favors a stripe domain configuration that fills the magnetic background to minimize the energy landscape. This stripe environment creates a spatial confinement around the scan region, effectively compressing the nucleated skyrmions into dense clusters. The scan-line spacing $\Delta$ was set to write a region filled with uniformly sized skyrmions, thereby minimizing topological defects arising from uneven neighboring sites. By positioning the scanning probe between adjacent skyrmions, we harness the competing repulsive forces to drive and align the skyrmions along a fixed writing direction\cite{ge2023constructing}. To quantify the ordering of our skyrmion lattices, we calculate the local orientational order parameter $\psi$\textsubscript{6}, defined as
\begin{equation}
    \psi_6(\mathbf{r}_j) = \frac{1}{N} \sum_{k=1}^{N} e^{-i6\theta_{jk}}
    \label{eq:placeholder}
\end{equation}
\\
for a particle at position $\mathbf{r}_j$ with $N$ neighbors and $\theta_{jk}$ as the orientational angle of the connecting vector from $\mathbf{r}_j$ to $\mathbf{r}_k$ with respect to an arbitrary fixed axis\cite{gruber2025real}. $\psi_6$ can then be used to indicate whether the skyrmions are in an arrangement that can be described as a liquid phase, hexatic phase, or solid phase\cite{klaui2020freezing}. We used a machine-learning-based pixel-wise classification to detect skyrmions by their size and morphology\cite{labrie2024machine} (see Supplementary Material S1). The obtained positions were used for every skyrmion to determine the local order parameter $\psi_6$, the orientational angle $\theta$, and its nearest neighbors, applying a Voronoi tessellation to identify the lattice defects. Any lattice site with a number of nearest neighbors different from six is classified as a topological defect. Skyrmions at the edge of the system and/or neighboring a stripe are neglected for the analysis, as their position produces artifacts in the Voronoi tessellation. The local orientational order parameter $|\psi_6|$ was calculated for every skyrmion, and the mean value $\langle|\psi_6|\rangle$ represents the overall lattice order.

\begin{figure}
\includegraphics{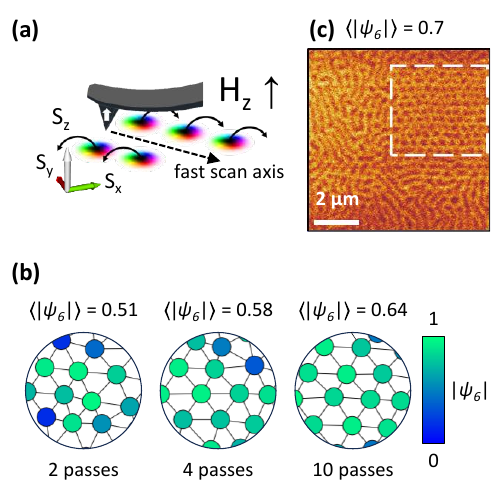}
\caption{\label{fig:epsart3} Improving the skyrmion lattice order. (a) Schematic of the local tip-induced repulsion due to antiparallel magnetization used for rearranging skyrmions along the writing direction. (b) Evolution of the mean lattice order parameter $\langle|\psi_6|\rangle$, achieved by rearranging skyrmions after 2, 4, and 10 writing passes. Each circle gives the local order parameter $|\psi_6|$ per skyrmion within the lattice. (c) Ordered skyrmion lattice with $\langle|\psi_6|\rangle$ = 0.7. The dashed white box indicates the region where skyrmions were nucleated and rearranged.}
\end{figure}

Figure 3b demonstrates the process of improving lattice order. The lattices were imaged in the “imaging mode” after writing to analyze the resulting degree of order. The skyrmion positions following each sequence, along with their corresponding $|\psi_6|$ values and connections to nearest neighbors, are mapped and presented. Initially, two writing passes along the 0$^{\circ}$ direction were applied to a labyrinth domain pattern, stabilizing a low-order skyrmion lattice with $\langle|\psi_6|\rangle$ = 0.51. As shown in the first panel of Figure 3b, the resulting lattice was slightly misaligned relative to the writing direction. Four additional writing passes improved the order to $\langle|\psi_6|\rangle$ = 0.58, aligning the skyrmions along the 0$^{\circ}$ direction. Further applying four more passes yielded hexagonal lattices in which most skyrmions had six nearest neighbors, indicating fewer topological defects and higher local order, reflected by increased $|\psi_6|$ values. By optimizing the number of writing passes, skyrmions with non-six neighbors were displaced to more optimal positions in the lattice, thereby improving the overall lattice ordering and leading to a higher $\langle|\psi_6|\rangle$. Using ten consecutive writing passes, a well-ordered lattice was situated within a stripe domain environment, as shown in Figure 3c. Ten skyrmions were arranged along the 0$^{\circ}$ direction, with the homogeneous stripe background acting as a spatial confinement boundary to prevent escape and compressed the skyrmions into a near solid phase configuration with $\langle|\psi_6|\rangle$ = 0.7. Note that skyrmions in our ten-repetition system exhibit negligible diffusion due to enhanced thermal stability arising from interlayer exchange coupling. As a result, no self-driven rearrangement occurs, and the improved ordering observed is a direct consequence of our active manipulation (see Supplementary Material S3).

\begin{figure}
\includegraphics{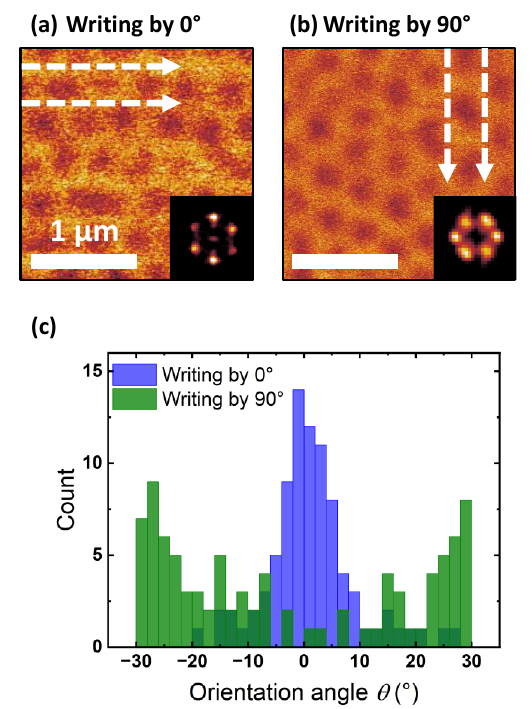}
\caption{\label{fig:epsar4} Rotating the skyrmion lattice orientation. (a) Hexagonal skyrmion lattices written along the horizontal (0$^{\circ}$) direction. (b) The same skyrmion lattice rotated by 90$^{\circ}$ after applying writing scans along the vertical direction. White arrows indicate the fast scan axis directions. (c) Statistical analysis of the orientational angle $\theta$ for skyrmion lattices written along the 0$^{\circ}$ and 90$^{\circ}$ directions.}
\end{figure}

Next, we demonstrate the ability to control the skyrmion lattice by tuning its orientation. We start by writing a lattice with skyrmions oriented along the 0$^{\circ}$ direction (horizontal) with some topological defects induced by uncut stripes. Then, another writing procedure was performed in the same area, but with the scan oriented along a 90$^{\circ}$ direction. The skyrmion lattices before and after the rotation are shown in Figures 4a and 4b, respectively. The subsequent skyrmion lattice orientation was rotated to align along the 90$^{\circ}$ direction. For a 2D hexagonal lattice with a sixfold rotational symmetry, rotations of integer multiples of 60$^{\circ}$ are identical. As such, a 30$^{\circ}$ shift in reciprocal space of the FFT in Figure 4b is an equivalent representation of a 90$^{\circ}$ rotation, confirming that the scanning field gradient of the tip induced a 90$^{\circ}$-oriented lattice. In 2D systems, the orientational quasi-long-range order can be reflected via the possibility of finding nearest neighboring skyrmions along a fixed axis relative to the orientational angle $\theta$\cite{halperin1978theory,nelson1979dislocation}. The orientational angle was calculated for each skyrmion, defined as $\theta = \arg(\psi_6)/6$, and analyzed to find skyrmions with similar orientation over the entire lattice.  4c presents the distributions of $\theta$ for lattices written along the 0$^{\circ}$ and 90$^{\circ}$ directions, respectively. Since our analysis considers a two-dimensional system with sixfold orientational symmetry, $\theta$ is calculated in the range from -30$^{\circ}$ to  +30$^{\circ}$, as other values can be mapped equivalently with a 60$^{\circ}$ rotation. For lattices written along the 0$^{\circ}$ direction, a distribution of $\theta$ centered around 0$^{\circ}$ was observed, confirming that most skyrmions have neighbors along the 0$^{\circ}$ direction. Writing scans at 90$^{\circ}$ shift the nearest neighbors toward the $\pm$ 30$^{\circ}$ angles, where we observed higher counts of skyrmions. A nearly zero count at $\theta$ = 0 demonstrates a 90$^{\circ}$-oriented lattice.

In conclusion, we develop and demonstrate a robust, minimally invasive technique for the local imaging and manipulation of skyrmion lattices using single-pass MFM. By decoupling topographic mapping from magnetic phase detection, our approach eliminates tip-induced perturbations that commonly compromise soft magnetic textures in the conventional two-pass scheme. This method enables nondestructive observation of soft magnetic textures and provides a useful tool to study skyrmion stability in pinning-dominated systems. Tapping scans were used to trigger reversible transitions from stripe domains to skyrmions by tuning the spacing $\Delta$ between adjacent scan lines, enabling reliable creation and deletion of energetically metastable textures on a homogeneous stripe background. When $\Delta$ $\approx$ $L$ in the demagnetized state, efficient stripe-to-skyrmion conversion with circular morphology and skyrmion lattices is facilitated. Further optimization of the scan parameters and the alignment of the initial domain pattern can result in a higher conversion yield. By operating near zero-field conditions, we leveraged the interactions originating from surrounding domains and tip-induced repulsion to align skyrmions and improve their arrangement along the writing direction. We calculate the local orientational order parameter $\psi_6$ from skyrmion positions to reveal a progressive improvement of the lattice order when passing the tip repeatedly in writing mode, inducing a transition from a disordered state to a lattice with long-range hexagonal order. In particular, we demonstrate that the lattice orientation can be freely rotated by changing the writing direction, as confirmed by the analysis of 2D-FFTs and the orientational angle $\theta$. Our approach establishes a pathway towards the tailored formation of skyrmionic crystals with artificial boundaries and tunable lattice attributes, which has not been previously possible, yet is crucial for applications, such as spin-wave devices, where the propagation of certain spin-wave modes depends on their direction relative to the skyrmion lattice.\\

\section*{Supplementary material}
See the supplementary material for details on the image classification used for lattice analysis, the correlation between $\Delta$ and $L$ for efficient stripe-to-skyrmion conversion, and the comparative analysis of ordering and orientation in spontaneously nucleated skyrmion lattices.\\

This project has received funding from the European Union’s Horizon Europe Programme Horizon under the Marie Skłodowska-Curie Actions (MSCA), Grant agreement No. 101119608 (TOPOCOM). It has also been supported by the Deutsche Forschungsgemeinschaft (DFG, German Research Foundation) – SPP 2137 Skyrmionics (project 403502522), TRR 173-268565370 Spin+X (Project A01, B02), and the European Research Council (ERC), Grant Agreement No. 856538 (3D MAGiC). M.K. and D.M. acknowledge support from the Norwegian Research Council through Grant No. 262633, Center of Excellence on Quantum Spintronics (QuSpin). E.M.J acknowledges the “Alexander von Humboldt Postdoctoral Fellowship”.

\section*{Author Declarations}

This article may be downloaded for personal use only. Any other use requires prior permission of the author and AIP 
Publishing. This article appeared in Appl. Phys. Lett. 127, 252407 (2025) and may be found at 
https://doi.org/10.1063/5.0301050.\\

\noindent
\textbf{Conflict of Interest}

The authors have no conflicts to disclose.\\

\noindent
\textbf{Author Contributions}

\noindent
\textbf{Duc Minh Tran}: Conceptualization (equal); Data curation (lead); Investigation (lead); Formal analysis (lead); Methodology (lead); Visualization (equal); Writing – original draft (lead); Writing – review \& editing (equal).  \textbf{Edoardo Mangini}: Investigation (equal); Formal analysis (equal); Methodology (equal). \textbf{Elizabeth M. Jefremovas}: Conceptualization (equal); Supervision (equal); Data curation (equal); Writing – review \& editing (equal). \textbf{Fabian Kammerbauer}: Methodology (equal); Writing – review \& editing (equal). \textbf{Dennis Meier}: Methodology (equal); Writing – review \& editing (equal). \textbf{Robert Fr{\"o}mter}: Supervision (equal); Visualization (equal); Writing – review \& editing (equal). \textbf{Mathias Kl{\"a}ui}: Conceptualization (lead); Resources (lead); Supervision (lead); Writing – review \& editing (equal).

\section*{Data Availability Statement}
The data supporting the findings in this study is publicly available via Zenodo at https://doi.org/10.5281/zenodo.17063771.

\bibliography{references}

\end{document}